\documentstyle[11pt,aaspp4]{article}

\newcommand{\dd}{{\rm d}}
\lefthead{MORI}
\righthead{THE GALACTIC DIFFUSE GAMMA-RAY SPECTRUM}	

\begin{document}

\slugcomment{\sl To appear in Astrophysical Journal 478 (March 20, 1997)}

\title{THE GALACTIC DIFFUSE GAMMA-RAY SPECTRUM FROM COSMIC-RAY PROTON
   INTERACTIONS}
\author{Masaki Mori\altaffilmark{1,2}}
\affil{NASA / Goddard Space Flight Center, Greenbelt, MD 20771}
\authoremail{m-mori3@ipc.miyakyo-u.ac.jp}
\altaffiltext{1}{On leave of absence from Department of Physics, Miyagi 
   University of Education, Sendai, Miyagi 980, Japan}
\altaffiltext{2}{E-mail address: {\tt m-mori3@ipc.miyakyo-u.ac.jp}}

\begin{abstract}
A new calculation of the Galactic diffuse gamma-ray spectrum from the decay 
of secondary particles produced by interactions of cosmic-ray protons with 
interstellar matter is presented. The calculation utilizes the modern Monte
Carlo event generators, {\sc Hadrin}, {\sc Fritiof} and {\sc Pythia}, 
which simulate 
high-energy proton-proton collisions and are widely used in studies of 
nuclear and particle physics, in addition to scaling calculation.
This study is motivated by the result on the Galactic diffuse
gamma-ray flux observed by the EGRET detector on the Compton Gamma-ray
Observatory, which indicates an excess above about 1 GeV of the observed 
intensity compared with a model prediction. 
The prediction is based on cosmic-ray interactions with interstellar matter,
in which secondary pion productions are treated by a simple model.
With the improved
interaction model used here, however, the diffuse gamma-ray flux agrees 
rather well with previous calculations within uncertainties, which mainly come
from the unobservable demodulated cosmic-ray spectrum in interstellar space.
As a possible solution to the excess flux, flatter spectra of cosmic-ray
protons have been tested and we found that the power-law spectrum with an 
index of about $-(2.4\sim2.5)$ gives a better fit to the EGRET data, 
though the spectrum is not explained completely.
\end{abstract}


\section{ Introduction }

Observations of diffuse gamma-ray emission from the Galactic plane 
give us some knowledge on Galactic cosmic-rays, interstellar medium and
interaction between them.
The observed features, most recently by the Energetic Gamma Ray Experiment
Telescope (EGRET) on the Compton Gamma-Ray Observatory, are described
fairly well by a model based on dynamic balance and realistic interstellar
matter and photon distributions (Hunter et al.~1996)\markcite{Hun96}.

However, the observed intensity exceeds the model prediction by as much
as 60\% for energies above about 1 GeV.
One of the possible explanation of this discrepancy is that the theory
of diffuse gamma-ray production may not be adequate at high energies: 
astrophysical gamma-rays above 1 GeV have never been measured before with 
high statistical accuracy (Hunter et al.~1996)\markcite{Hun96}.

There are three basic components for the production of Galactic diffuse 
gamma-rays: nuclear interactions between cosmic-rays and matter, 
bremsstrahlung collisions between electrons and matter, and inverse 
Compton scattering of electrons with low-energy photons.
Above about 200 MeV, the first component, more specifically the gamma-ray 
production from the decay of neutral pions produced in cosmic-ray
(protons and nuclei) collisions with interstellar matter, is known to be 
the dominant one.
Previous works on this component used isobaric models and scaling models
for the nuclear interaction: see Dermer (1986a)\markcite{Der86a} for detail.
These studies, however, were focused on lower energies.
This is natural since the gamma-ray spectrum from pion decay has a peak
around 70 MeV and drops rapidly toward higher energies.
In order to compare the observed high-energy diffuse gamma-ray emission
with model predictions, however, it is necessary to use more detailed models
which describe high-energy proton-proton ($p$-$p$) collisions more accurately.
Here we use Monte Carlo event generators that are commonly used in the 
analysis of high-energy physics experiments to simulate such collisions.
We apply these results for the calculation of the Galactic diffuse gamma-ray 
flux from cosmic-ray interactions up to $10^7$ GeV and compare it with 
previous studies and data from the EGRET detector.

\section{ Gamma-ray production in cosmic-ray interaction }

The gamma-ray production source function from nuclear interaction, or the 
gamma-ray spectrum resulting from pion decay in $p$-Hydrogen atom 
collisions, is given by Stecker (1970)\markcite{Ste70}
\begin{eqnarray}
q(E_\gamma) &=& 4\pi n_H \int_{E_\gamma+m_\pi^2/4E_\gamma}^\infty \dd E_\pi
\cdot2\int_{T_p^{\rm min}(T_\pi)}^\infty \dd T_p 
{j_p(T_p) \over \sqrt{E_\pi^2-m_\pi^2}}
{\dd\sigma(T_p,T_\pi) \over {\dd T_\pi}} \cr
&=& 4\pi n_H \int_{T_p^{\rm min}}^\infty \dd T_p j_p(T_p)
 \langle\zeta\sigma_\pi(T_p)\rangle \cr
&& \times \int_{E_\gamma+m_\pi^2/4E_\gamma}^\infty \dd E_\pi 
{{2 \dd N(T_p,T_\pi) / \dd T_\pi} \over \sqrt{E_\pi^2-m_\pi^2}} 
\label{eq:srcfn}
\end{eqnarray}
where the cosmic-ray proton flux is denoted by $j_p(T_p)$ 
[cm$^{-2}$s$^{-1}$sr$^{-1}$GeV$^{-1}$], $\dd\sigma(T_p,T_\pi)/\dd T_\pi$
is the differential cross section for the production of a $\pi^0$ 
with kinetic energy $T_\pi$
in the Galactic (rest) system due to a collision of a cosmic-ray proton of
kinetic energy $T_p$ with a H atom at rest, $E_\pi$ is the total pion energy 
and $m_\pi$ is its mass, and $n_H$ is the atomic hydrogen density.
$T_p^{\rm min}(T_\pi)$ is the minimum proton kinetic energy that contributes
to the production of a pion with energy $T_\pi$.
It can be calculated from the kinematics easily.
Following Dermer\markcite{Der86a} (1986a), we write 
$\dd\sigma(T_p,T_\pi)/\dd T_\pi =
 \langle\zeta\sigma_\pi(T_p)\rangle \dd N(T_p,T_\pi) / \dd T_\pi$, where
$\langle\zeta\sigma_\pi(T_p)\rangle$ is the inclusive cross section
for the reaction $p+p\rightarrow \pi^0+{\rm anything}$ 
($\zeta$ is the $\pi^0$ multiplicity)
and $\dd N(T_p,T_\pi) / \dd T_\pi$ is the normalized $\pi^0$ spectrum:
\begin{equation}
\int_0^\infty \dd T_\pi {\dd N(T_p,T_\pi) \over \dd T_\pi} = 1.
\end{equation}

\subsection{ Secondary pion production } \label{sec:2nd}

We studied the three Monte Carlo programs {\sc Hadrin} 
(H\"{a}nssgen and Ranft 1986)\markcite{Han86}, {\sc Pythia} 
(St\"{o}strand 1994)\markcite{Sjo94},
and {\sc Fritiof} (Pi 1992)\markcite{Pi92},
in comparison with the scaling model of
the $\pi^0$ production cross section by $p$-$p$ interaction given in an
analytical form by Stephens and Badhwar\markcite{Ste81} (1981). 
This scaling model was used in some of the previous calculations
(Stephens and Badhwar\markcite{Ste81} 1981; Dermer\markcite{Der86a} 1986a).

The first program, {\sc Hadrin}, focuses on a threshold and resonance 
behavior of inelastic hadron-nucleon interactions and uses tabulated 
cross sections for many possible reaction channels based on experimental data.
It focuses on describing the nuclear collision at laboratory energies below 
5 GeV.
Fig.~\ref{fig:tp97cms} shows the comparison of the simulated $\pi^0$
kinetic energy spectrum with experimental data and analytic calculations
based on the isobar model (Stecker\markcite{Ste70} 1970) and the scaling
model (see also Dermer\markcite{Der86a} 1986a) in the center-of-mass system 
(CMS).
Also shown in Fig.~\ref{fig:tp97ls} is the same data but in laboratory 
system (LS).
One can see both {\sc Hadrin} and the isobar model calculations show good 
agreement with experimental data 
within limited statistical accuracy of the data.

\placefigure{fig:tp97cms}
\placefigure{fig:tp97ls}

The program called {\sc Pythia}\footnote{
We used {\sc Pythia} version 5.718 with {\sc Jetset} version 7.408.}
focuses on high-energy $p$-$p$ colliders
and its performance is reported in detail (Sj\"{o}strand and van 
Zijl\markcite{Sjo87} 1987).
This is based on a string-fragmentation model, but incorporates
low transverse momentum ($p_T$) interactions, which is our main concern.
Its usefulness for fixed target experiments, the situation similar
to ours, is not fully studied, however.
The other program, {\sc Fritiof},\footnote{
We used {\sc Fritiof} version 7.02 with {\sc Ariadne} version 4.02R and 
{\sc Jetset} version 7.3.}
implements a model for low-$p_T$ hadron-hadron,
hadron-nucleus and nucleus-nucleus reactions which treats a hadron
as a string-like object with its color force field stretching like
a vortex line.
A detailed description of the model and its comparison with experimental 
data are reported (Andersson, Gustafson and Pi\markcite{And93} 1993).

Here we compare their predictions with the $p$-$p$ collision data:
due to the experimental difficulty in obtaining $\pi^0$ spectra in
high-multiplicities, we use gamma-ray spectra.

Fig.~\ref{fig:rapid} shows the CMS rapidity 
($y^* \equiv {1 \over 2}\ln{{E^*+p_L^*}\over{E^*-p_L^*}}$, where $E^*$ is 
the pion energy in CMS
and $p_L^*$ the longitudinal momentum in CMS) distributions for beam
momenta of 12.4, 205, and 300 GeV/c.
Data are obtained in bubble chamber experiments at Argonne
(Jaeger et al.\markcite{Jae75a} 1975a) and Fermilab
(Jaeger et al.\markcite{Jae75b} 1975b; Sheng et al.\markcite{She75} 1975).
Predictions and data are normalized to match the total cross sections
given by the fitted formula given by Dermer\markcite{Der86b} (1986b).
The scaling calculation has been executed as a combination of the analytic 
Stephens and Badhwar\markcite{Bad81} (1981) model and a Monte Carlo program 
{\sc Decay} (H\"{a}nssgen and Ritter\markcite{Han84} 1984) to treat 
$\pi^0$ decays.
One sees a general agreement of data with Monte Carlo predictions,
but some overprediction is seen in the forward and backward 
($|y^*|\gtrsim2$) directions at high energies.
This tendency is more evident for {\sc Pythia}: it may partly due to its main
intention to describe high transverse momentum ($p_T$) phenomena, while 
{\sc Fritiof} is a model to describe low $p_T$ hadronic reactions.
We do not use {\sc Pythia} to calculate the gamma-ray flux since the forward
region plays an importrant role at high energies (see below).

Fig.~\ref{fig:invcrs} shows the comparison of the model prediction of
the invariant cross section of $\pi^0$ production with
experimental data at ISR.
The differential cross sections of gamma-rays based on Monte Carlo programs 
have been converted to invariant cross sections of $\pi^0$'s using
Sternheimer relation (Sternheimer\markcite{Ste55} 1955)
following the strategy given in Stephens and Badhwar\markcite{Ste81}
(1981).\footnote{
We have used the correct formula: $E(\dd^3\sigma/\dd p^3) = 
-(1/2)(\partial/\partial p)(\dd^2\sigma
/(\dd p\dd\Omega)) \approx (1/2)(B+2C)(\dd^2\sigma/(\dd p\dd\Omega))$ if
$\dd^2\sigma/(\dd p\dd\Omega)\propto\exp(-Bp-Cp^2)$.}
Monte Carlo calculations agree well with the scaling calculations given
by Stephens and Badhwar, but both
of them overpredict the cross sections at higher rapidities, as
was seen in the previous figure (see
the discussion of Stephens and Badhwar about the significance of this
discrepancy). This results in systematic uncertainties of the final
gamma-ray flux at higher energies ($E_\gamma\gtrsim100$ GeV).

\placefigure{fig:rapid}
\placefigure{fig:invcrs}

\subsection{ Inclusive cross section }

Measured inclusive cross sections, $\langle\zeta\sigma_\pi(T_p)\rangle$,
for the reaction $p+p\rightarrow \pi^0+{\rm anything}$ are
well fitted by the analytic parameterization given by Dermer\markcite{Der86b}
(1986b) for $T_p \le 1$ TeV.
In order to extend the calculation to higher energies, inclusive cross
sections have been computed with a help of Monte Carlo simulators.
In order to match the experimental data, inclusive cross sections
of $\pi^0$ production are computed from inclusive cross sections of
gamma-ray production assuming all gamma-rays are from $\pi^0$ decays.
Fig.~\ref{fig:picrs} compares experimental results with the Dermer's 
parameterization and the {\sc Pythia} and {\sc Fritiof} prediction.
Total $p$-$p$ cross sections are calculated by a parameterization inspired by 
Regge theory (Donnachie and Landshoff\markcite{Don92} 1992) in {\sc Pythia}
and by a Block-Cahn fit (Block and Cahn\markcite{Blo87} 1987) in {\sc Fritiof}.
We adopt Dermer's parameterization below 1 TeV/c and use 
$163(s/1876\,{\rm GeV}^2)^{0.21}$ mb as a smooth fit to the {\sc Pythia} prediction
above 1 TeV/c, where $s=2m_p(T_p+2m_p)$ with $m_p$ the proton mass, since
it is based on newer fit to the total cross section and charged particle
multiplicity distributions are described well up to $\sqrt{s}=900$ GeV
(Sj\"{o}strand and van Zijl\markcite{Sjo87} 1987).

\placefigure{fig:picrs}

\subsection{ Cosmic-ray proton flux }

Galactic diffuse gamma-rays are mostly produced in the central region of the
Galaxy and we should use the cosmic-ray spectrum there: however, since
our knowledge is limited within the local solar neighborhood, the best we can
do is to use the local cosmic-ray spectrum which is estimated by using
appropriate assumptions from the solar-modulated spectrum actually observed.
Fig.~\ref{fig:crspec} summarizes the observations and demodulated spectra
calculated by different authors.
There is a factor of about 5 uncertainty in the proton spectrum at $T_p=1$ 
GeV, but as we will see these low energy protons do not contribute much to 
the diffuse flux, as has been pointed out before (e.g.~Stephens and 
Badhwar\markcite{Ste81} 1981).
In order to quantify the spectrum uncertainty, we will calculate the
diffuse gamma-ray flux for three extreme cases: the ``maximal'' (``minimal'')
flux corresponds to the upper (lower) envelope of calculated demodulated
spectra and connects to $E^{-2.75}$ spectra above 100 GeV,
and the ``median'' flux takes the form of
\begin{equation}
\begin{array}{lll}
J_p(T_p) &=& \left\{\begin{array}{ll}
                     1.67 p_p^{-2.7}\left[1+\left({2.5\,{\rm GeV/c} \over p_p}
                     \right)^2\right]^{-1/2} & (E_p\le100\,{\rm GeV}) \cr
                     6.65\times10^{-6}\left({E_p \over 100\,{\rm GeV}}
                     \right)^{-2.75} & (E_p>100\,{\rm GeV})
                    \end{array} \right.
\end{array}
\end{equation}
where $E_p=T_p+m_p$, $p_p=\sqrt{E_p^2-m_p^2}$ and the unit is
cm$^{-2}$s$^{-1}$sr$^{-1}$. These are inspired by the parameterization
on the demodulated spectrum given by Ormes and Protheroe\markcite{Orm83} 
(1983) for $E_p\le100$ GeV and
the fit by Honda et al.\markcite{Hon95} (1995) for higher energies.
The cosmic-ray proton fluxes used in this work are shown in 
Fig.~\ref{fig:crspecMM}.
Notice that these fluxes are {\it extremes} and do not represent {\it typical}
errors.
Since the direct observation of cosmic-ray composition is limited to energies
less than $10^6$ GeV/nucleus, and there may be a possible steepening of
the proton flux above 40 TeV (Asakimori et al.\markcite{Asa93} 1993),
flux of diffuse gamma-rays above about $10^4$ GeV suffers large uncertainty.

\placefigure{fig:crspec}
\placefigure{fig:crspecMM}

\subsection{ Effect of heavier nuclei }

The effect of heavier nuclei than proton in cosmic-ray and the interstellar
matter was estimated by different authors.
It has been treated as a constant multiplification factor independent of
energy.
This is valid for the calculation of diffuse gamma-rays in the GeV region,
since the proton/helium ratio in cosmic-rays seems to be constant in
the 10$\sim$100 GeV range.
However, recent observations show different spectral indices between protons
and other heavier nuclei (see Biermann, Gaisser and Stanev\markcite{Bie94} 
1994; Wiebel-Sooth, Bierman and Meyer\markcite{Wie95} 1995 
for summary).
This difference may be explained by a model in which protons and heavier 
nuclei come from different kinds of sources (Biermann, Gaisser and 
Stanev\markcite{Bie94} 1994).
Therefore we derive an energy dependence of this nuclear enhancement factor,
$\epsilon^M$, assuming $E^{-2.75}$ spectrum for protons and $E^{-2.63}$
spectra for heavier nuclei (Wiebel-Sooth, Bierman and Meyer\markcite{Wie95}
1995).
Following Gaisser and Shaefer\markcite{Gai92} (1992), $\epsilon^M$ can be 
written as
\begin{equation}
\epsilon^M = \sum_i^{\rm CR} \left({n_i \over n_p}\right)_{\rm CR}
   {1 \over 2}\left[w_{ip}{\sigma_{ip} \over \sigma_{pp}}
    + {\sigma_{i\alpha} \over \sigma_{pp}}
      \left({n_\alpha \over n_{\rm H}}\right)_{\rm ISM}w_{i\alpha}\right]\,,
\end{equation}
dropping a factor which is related to propagation (only necessary for the
case of antiprotons discussed in Gaisser and Shaefer\markcite{Gai92} (1992), 
where $(n_i/n_p)_{\rm CR}$
is the ratio of numbers of nuclei of type $i$ to protons (the subscript CR
indicates the quantity in cosmic-rays), $w_{ip}(w_{i\alpha})$
the total number of wounded nucleons in a collision between a nucleus $i$ and
a proton ($\alpha$), $\sigma_{ip}(\sigma_{i\alpha})$ the total inelastic cross
section for a collision between a nucleus $i$ and a proton ($\alpha$), and
$(n_\alpha/n_{\rm H})_{\rm ISM}$ is the ratio of helium to hydrogen in
interstellar matter (ISM).
The ISM is assumed to be a mixture of 93\% hydrogen and 7\% helium after
Garcia-Munoz et al.\markcite{Gar87} (1987).
We assume
\begin{equation}
\begin{array}{lll}
{\displaystyle \left({n_i \over n_p}(T_p)\right)_{\rm CR}} &=&
  {\displaystyle \left({n_i \over n_p}({\rm GeV})\right)_{\rm CR}}
    \times\left\{\begin{array}{ll} 1 
            & (i=p\,\,\,{\rm or}\,\,\,T_p\le100\,{\rm GeV})\cr
             \left({T_p \over 100\,{\rm GeV}}\right)^{0.12}
            & ({\rm otherwise}).\end{array}
          \right.
\end{array}
\end{equation}
where $(n_i/n_p({\rm GeV}))_{\rm CR}$ is the cosmic-ray
abundance in GeV range and is tabulated in Gaisser and Shaefer\markcite{Gai92}
(1992).
The result is plotted in Fig.~\ref{fig:epsM} which is to be compared with
previous estimates: 1.5 by Cavallo and Gould\markcite{Cav71} (1971), 
$1.6\pm0.1$ by Stephens and Badhwar\markcite{Ste81} (1981), 1.45 by 
Dermer\markcite{Der86a} (1986a).
$\epsilon^M$ takes a value of 1.52 for $T_p<100$ GeV, as given in
Gaisser and Shaefer\markcite{Gai92} (1992).
Notice that in the above expression the relativistic rise of interaction
cross sections is cancelled out.
The differences may be partly attributed to a larger multiplicity
enhancement factor used in Stephens and Badhwar\markcite{Ste81} (1981) 
and a lower helium/proton ratio used in Dermer\markcite{Der86a} (1986a).

\placefigure{fig:epsM}

\subsection{ Monte Carlo calculation }

Monte Carlo simulators can produce gamma-rays as final products instead of
$\pi^0$'s, including the decay kinematics.
Also other secondaries, such as neutral kaons which may yield gamma-rays,
are included, although their contribution is minor.
Thus we generated histograms for gamma-rays, not for $\pi^0$'s, 
for various proton energies with logarithmic binning (1 decade = 10 bins).
In this case the equation (\ref{eq:srcfn}) can be written in a summation 
form of
\begin{equation}
q(E_{\gamma,j})\cdot\Delta E_\gamma = 
4\pi n_H \sum_i \Delta T_{p,i}\,j_p(T_{p,i})
\langle\zeta\sigma_\pi(T_{p,i})\rangle f(T_{p,i};E_{\gamma,j}),
\end{equation}
or, using $\Delta E_{\gamma,j}=E_{\gamma,j}\Delta$ and
$\Delta T_{p,i}=T_{p,i}\Delta$ where $\Delta=10^{0.05}-10^{-0.05}$,
\begin{equation}
q(E_{\gamma,j}) = 4\pi n_H \sum_i j_p(T_{p,i})
\langle\zeta\sigma_\pi(T_{p,i})\rangle f(T_{p,i})\, T_{p,i}/E_{\gamma,j}
\end{equation}
with $f(T_{p,i};E_{\gamma,j})$ the normalized value of the histograms 
at $E_{\gamma,j}$ ($\sum_i f(T_{p,i};E_{\gamma,j})
=2$ here to incorporate the fact that 2$\gamma$'s are produced by a $\pi^0$).
In the range of energies
3 GeV $<T_p<$ 12.5 GeV no complete experimental data are available.
Therefore we decided to use {\sc Hadrin} for $T_p\le8$ GeV, the scaling 
model for $T_p\ge3$ GeV, and {\sc Fritiof} for $T_p>10$ GeV: 
we joined models with a linear connection
in the overlapping energy regions (3--8 GeV) to obtain a smooth curve.
Above 10 GeV results from two models will be compared.
For Monte Carlo programs 
80,000$\sim$1,000 events were generated for incident proton energies of
$T_p=10^{-0.5},10^{-0.4},$ $10^{-0.3},\ldots,10^{6.9},10^{7}$ GeV, 
respectively, depending on energy in order to have enough statistical 
accuracy while saving computing time.

\section{ Results }

Fig.~\ref{fig:spec_d} shows our results on the emissivity of gamma-rays from
the interaction of cosmic-rays with unit density of atomic hydrogen for
two interaction models for $T_p\ge10$ GeV, the scaling model and {\sc Fritiof}.
The statistical errors due to Monte Carlo calculations are (typically): 4, 
0.6, 0.8, 2, and 3\% at $E_\gamma=$ 0.01, 0.1, 1, 10, and 100 GeV, 
respectively.
For the cosmic-ray proton flux, we assumed three cases: ``median'',
``maximum'', and ``minimum''.
(For {\sc Fritiof} we only show the result with the ``median'' flux.)
The ``maximal'' (``minimal'') proton flux gives about 50\% higher (lower) 
gamma-ray flux at around 0.1 GeV and about 20\% at around 10 GeV compared with
that with the ``median'' flux.
Remembering that these are {\it extremes}, we may say that the {\it standard}
uncertainty of the calculated flux derived from the uncertainty of cosmic-ray 
proton flux is around 20 to 30\%.
The calculated emissivity is 34\% higher at 10 GeV when we use {\sc Fritiof}
compared with the scaling model: this may be a measure of systematic 
uncertainty related to the interaction model of the present predicition.

Table \ref{table:emis} summarizes the calculated gamma-ray emissivities 
for three cases of the cosmic-ray proton flux in differential and integral
form with the scaling model.

\placefigure{fig:spec_d}

\placetable{table:emis}

\begin{deluxetable}{cccccccc}
\tablecolumns{8}
\small
\tablecaption{Gamma-ray emissivity due to secondary particle
production in collisions of cosmic-rays with interstellar matter.
\label{table:emis}}
\tablewidth{0pt}
\tablehead{
\colhead{Energy} & 
\multicolumn{3}{c}{Differential rate in} & &
\multicolumn{3}{c}{Integral rate in}\\
\colhead{in} & \multicolumn{3}{c}{photon s$^{-1}$GeV$^{-1}n_{\rm H}^{-1}$} &&
\multicolumn{3}{c}{photon s$^{-1}n_{\rm H}^{-1}$} \\
\cline{2-4}\cline{6-8}
\colhead{GeV} & \colhead{minimum} &  \colhead{median} & \colhead{maximum} &
\colhead{}    & \colhead{minimum} &  \colhead{median} & \colhead{maximum} }
\startdata
1.000E-02& 5.872E-26& 9.960E-26& 1.380E-25&& 9.335E-26& 1.646E-25& 2.385E-25\nl
1.585E-02& 1.133E-25& 2.019E-25& 2.924E-25&& 9.298E-26& 1.639E-25& 2.376E-25\nl
2.512E-02& 1.894E-25& 3.540E-25& 5.358E-25&& 9.188E-26& 1.619E-25& 2.347E-25\nl
3.981E-02& 2.478E-25& 4.726E-25& 7.298E-25&& 8.915E-26& 1.568E-25& 2.268E-25\nl
6.310E-02& 2.721E-25& 5.223E-25& 8.124E-25&& 8.384E-26& 1.466E-25& 2.112E-25\nl
1.000E-01& 2.581E-25& 4.937E-25& 7.652E-25&& 7.495E-26& 1.296E-25& 1.847E-25\nl
1.585E-01& 2.088E-25& 3.937E-25& 6.006E-25&& 6.206E-26& 1.050E-25& 1.467E-25\nl
2.512E-01& 1.333E-25& 2.413E-25& 3.549E-25&& 4.651E-26& 7.594E-26& 1.027E-25\nl
3.981E-01& 6.908E-26& 1.175E-25& 1.630E-25&& 3.164E-26& 4.940E-26& 6.416E-26\nl
6.310E-01& 3.206E-26& 5.116E-26& 6.684E-26&& 1.977E-26& 2.949E-26& 3.690E-26\nl
1.000E+00& 1.301E-26& 1.961E-26& 2.447E-26&& 1.131E-26& 1.618E-26& 1.971E-26\nl
1.585E+00& 4.765E-27& 6.861E-27& 8.314E-27&& 5.986E-27& 8.229E-27& 9.847E-27\nl
2.512E+00& 1.546E-27& 2.125E-27& 2.524E-27&& 2.982E-27& 3.946E-27& 4.681E-27\nl
3.981E+00& 4.812E-28& 6.348E-28& 7.494E-28&& 1.455E-27& 1.866E-27& 2.216E-27\nl
6.310E+00& 1.497E-28& 1.910E-28& 2.266E-28&& 7.027E-28& 8.805E-28& 1.052E-27\nl
1.000E+01& 4.579E-29& 5.698E-29& 6.822E-29&& 3.324E-28& 4.108E-28& 4.934E-28\nl
1.585E+01& 1.361E-29& 1.673E-29& 2.016E-29&& 1.539E-28& 1.894E-28& 2.280E-28\nl
2.512E+01& 3.952E-30& 4.854E-30& 5.852E-30&& 7.028E-29& 8.672E-29& 1.041E-28\nl
3.981E+01& 1.133E-30& 1.398E-30& 1.678E-30&& 3.192E-29& 3.955E-29& 4.731E-29\nl
\tablebreak
6.310E+01& 3.238E-31& 4.015E-31& 4.799E-31&& 1.451E-29& 1.804E-29& 2.151E-29\nl
1.000E+02& 9.277E-32& 1.153E-31& 1.375E-31&& 6.625E-30& 8.248E-30& 9.820E-30\nl
1.585E+02& 2.669E-32& 3.324E-32& 3.956E-32&& 3.039E-30& 3.788E-30& 4.505E-30\nl
2.512E+02& 7.721E-33& 9.626E-33& 1.145E-32&& 1.403E-30& 1.750E-30& 2.079E-30\nl
3.981E+02& 2.248E-33& 2.805E-33& 3.333E-33&& 6.514E-31& 8.129E-31& 9.656E-31\nl
6.310E+02& 6.590E-34& 8.224E-34& 9.768E-34&& 3.042E-31& 3.797E-31& 4.509E-31\nl
1.000E+03& 1.941E-34& 2.423E-34& 2.878E-34&& 1.427E-31& 1.782E-31& 2.115E-31\nl
1.585E+03& 5.747E-35& 7.175E-35& 8.518E-35&& 6.724E-32& 8.395E-32& 9.967E-32\nl
2.512E+03& 1.708E-35& 2.133E-35& 2.532E-35&& 3.180E-32& 3.971E-32& 4.714E-32\nl
3.981E+03& 5.098E-36& 6.365E-36& 7.556E-36&& 1.510E-32& 1.885E-32& 2.238E-32\nl
6.310E+03& 1.527E-36& 1.906E-36& 2.263E-36&& 7.188E-33& 8.976E-33& 1.065E-32\nl
1.000E+04& 4.587E-37& 5.727E-37& 6.798E-37&& 3.432E-33& 4.285E-33& 5.086E-33\nl
1.585E+04& 1.383E-37& 1.726E-37& 2.049E-37&& 1.642E-33& 2.050E-33& 2.434E-33\nl
2.512E+04& 4.181E-38& 5.221E-38& 6.197E-38&& 7.863E-34& 9.819E-34& 1.165E-33\nl
3.981E+04& 1.269E-38& 1.585E-38& 1.881E-38&& 3.759E-34& 4.694E-34& 5.572E-34\nl
6.310E+04& 3.856E-39& 4.816E-39& 5.716E-39&& 1.784E-34& 2.227E-34& 2.644E-34\nl
1.000E+05& 1.153E-39& 1.440E-39& 1.709E-39&& 8.340E-35& 1.042E-34& 1.236E-34\nl
\enddata
\tablecomments{Figures to be read as 5.782E-26 = $5.782\times10^{-26}$, etc.
``Minimum'', ``median'' and ``maximum'' denote the models of the cosmic-ray
spectrum (see text). The scaling model is used for $T_p\ge10$ GeV.}
\end{deluxetable}

At this stage it may be worth showing the contributions of cosmic-ray protons 
of various energies to gamma-rays of some specific energies 
in Fig.~\ref{fig:fixedspec}.
From this figure one can see most diffuse gamma-rays below about 300 MeV
of nuclear origin come from protons with kinetic energies of a few to
several GeV, where no experimental results for secondary pion production
are available and where there is a large uncertainty in the cosmic-ray 
proton flux.

\placefigure{fig:fixedspec}

The results obtained here are compared with previous calculations 
(Cavallo and Gould\markcite{Cav71} 1971, Stecker\markcite{Ste79} 1979,
Stephens and Badhwar\markcite{Ste81} 1981, Dermer\markcite{Der86a} 1986a) 
in Figs.~\ref{fig:spec_cmp} and 
\ref{fig:spec_i}, where in the latter they are plotted in integrated form:
$q(>E_\gamma)=\int_{E_\gamma}^\infty q(E_\gamma')\dd E_\gamma'$.
The main difference between authors may come from the assumed demodulated
spectrum of the cosmic-ray protons.
(See Fig.~1 of Dermer\markcite{Der86a} (1986a) for comparison of assumed
spectra.)

\placefigure{fig:spec_cmp}
\placefigure{fig:spec_i}

We can extend our calculation to higher energy gamma-rays and the results
are presented in Fig.~\ref{fig:specHE} with previous calculations
(Stecker\markcite{Ste79} 1979, Berezinsky et al.\markcite{Ber93a} 1993,
Chardonnet et al.\markcite{Cha95} 1995).
For this plot we show calculation only with the scaling model: {\sc Fritiof}
causes fatal errors at $\gtrsim10$ TeV.
Since the mean interaction length of gamma-rays for photon-photon collisions
with the cosmic microwave background radiation becomes comparable to the 
distance to the Galactic center from Earth above several 100 TeV
(Protheroe\markcite{Pro86} 1986), gamma-ray flux at higher energies than 
this are greatly reduced.
Thus we stop our calculation at 100 TeV.
The proton fluxes used in other calculations are: $2.35E_p^{-2.67}$ 
(Stecker) and $1.75E_p^{-2.73}$ (Chardonnet et al.), 
where $E_p$ is in GeV and the unit is cm$^{-2}$s$^{-1}$sr$^{-1}$GeV$^{-1}$.
Berezinsky et al.\ assumed the {\it total} cosmic-ray spectrum to be 
$1.59E_p^{-2.73}$ cm$^{-2}$s$^{-1}$sr$^{-1}$(GeV/nucleon)$^{-1}$.
Also shown is the effect of energy-dependent nuclear enhancement factor
($\epsilon^M$), which raise the expected flux by about 10\% at 1 TeV.
The flux given in Berezinsky et al.\ is smaller than 
others, partly due to the assumed cosmic-ray spectrum.

\placefigure{fig:specHE}

In Table \ref{table:emis_cmp} the integral gamma-ray emissivities
calculated by different authors are summarized.

\placetable{table:emis_cmp}

\begin{deluxetable}{llcccc}
\footnotesize
\tablecaption{Integrated Galactic gamma-ray emissivity from cosmic-ray 
interactions with interstellar matter. \label{table:emis_cmp}}
\tablewidth{0pt}
\tablehead{
\colhead{Reference} & \colhead{Model} & 
\multicolumn{4}{c}{$q(>E_\gamma)$\tablenotemark{a}} \\
\cline{3-6}
\colhead{} & \colhead{} &  \colhead{0} & \colhead{0.1 GeV} & \colhead{1 GeV}
& \colhead{1 TeV} }
\startdata
Cavallo and Gould\markcite{Cav71} 1971&
 Isobar + Phase space & 2.4 & 1.8 & 0.13 & \nodata \nl
Stephens and Badhwar\markcite{Ste81} 1981&
 Scaling & 1.92--2.34 & 1.37--1.63 & 0.115--0.116 &\nodata \nl
Dermer\markcite{Der86a} 1986a&
 Isobar + Scaling & 2.02 & 1.53 & 0.159 & \nodata \nl
Stecker\markcite{Ste88} 1988\tablenotemark{b}&
 Isobar + Scaling & 1.97 & 1.59 & 0.209 & \nodata \nl
Berezinsky et al.\markcite{Ber93a} 1993&
 SIBYLL & \nodata & \nodata & \nodata & $0.76\times10^{-6}$ \nl
Chardonnet et al.\markcite{Cha95} 1995&
 {\sc Pythia} & \nodata & \nodata & \nodata & $1.5\times10^{-6}$ \nl
This work\tablenotemark{c} &
 {\sc Hadrin} + Scaling  & 1.65 & 1.30 & 0.162 & $1.78\times10^{-6}$ \nl
 & &(0.94--2.39) &(0.75--1.85) &(0.113--0.197) 
 & ((1.43--2.12)$\times10^{-6}$) \nl
\enddata

\tablenotetext{a}{Units: $10^{-25}$ s$^{-1}$ per hydrogen atom.}
\tablenotetext{b}{Computed using the parameterization given in 
Bertsch et al.\markcite{Ber93} (1993).}
\tablenotetext{c}{The ``median'' proton flux is used. (Emissivities with
the ``minimal'' and ``maximal'' proton fluxes are shown in parentheses.)}
\end{deluxetable}

\section{ Discussion }

Now we compare our results with the EGRET data (Hunter et al.\markcite{Hun96}
1996).

Fig.~\ref{fig:flux_cmp} shows the Galactic diffuse gamma-ray flux in a 
differential form, multiplied by $E_\gamma^2$ to make the differences more
visible.
The dotted line shows the expectation as the sum of three components of 
diffuse gamma-rays: cosmic-ray nuclear interactions (which is the main 
concern of this paper and is dominant above about 200 MeV), electron 
bremsstrahlung, and inverse Compton. 
The components were calculated by the model developed by Hunter
et al.\markcite{Hun96} (1996), which incorporates the detailed distribution of
interstellar matter, and are averaged over the Galactic center region,
$300^\circ < \ell < 60^\circ$, $|b|\le10^\circ$.
In their model they used the gamma-ray emissivity from cosmic-ray interactions
given by Stecker\markcite{Ste88} (1988).
Solid and dashed lines are the expected fluxes in which the 
gamma-ray emissivity is replaced with the ones computed in this work: solid
line assumes the ``median'' proton flux and dashed lines assume the 
``minimal'' and ``maximal'' fluxes.
None of the models mentioned above seems to yield enough gamma-rays above
1 GeV: thus we may conclude that even the detailed model studied here cannot
explain the excess of Galactic diffuse gamma-ray flux above 1 GeV.

As one of the possible solutions to the problem, we tried to fit the 
cosmic-ray proton spectrum from the EGRET data with a single power-law 
spectrum in total energy, $J_p(E_p) = a E_p^{-b}$, 
and have computed the $\chi^2$'s in the $a$-$b$ plane
and searched the minimum.

The results are $a=1.48^{+0.17}_{-0.20}$ 
cm$^{-2}$s$^{-1}$sr$^{-1}$GeV$^{-1}$ and $b=2.45^{+0.03}_{-0.04}$
when we use the scaling model for $T_p\ge10$ GeV
($a=1.67^{+0.23}_{-0.20}$ cm$^{-2}$s$^{-1}$sr$^{-1}$GeV$^{-1}$ and 
$b=2.51\pm0.04$ with {\sc Fritiof} for $T_p\ge10$ GeV).
The expected flux calculated from this best fit is plotted in 
Fig.~\ref{fig:flux_cmp} as a dotdashed line.
This spectrum gives a better fit above 1 GeV, but predicts higher
gamma-ray flux than observed around several 100 MeV.

\placefigure{fig:flux_cmp}

However, this flatter proton spectrum is not consistent with the direct
observations above 10 GeV (see Fig.~\ref{fig:crspecMM}).
Therefore one may be tempted to suggest a possibility
that the cosmic-ray spectrum is flatter in the Galactic center region 
than the one in the local (solar neighborhood) region.
It is interesting to note that the power-law index obtained here is similar 
to that of the cosmic-ray source spectrum predicted by the cosmic-ray 
reacceleration theory (Ptuskin\markcite{Ptu95} 1995 and references therein):
in this theory the source spectrum is proportional to $R^{-\gamma_s}$,
where $\gamma_s=2.3\sim2.4$ (where $R$ is the rigidity) and the escape 
length to $R^{-1/3}$ to match the observed spectrum at all energies.

In any case, more detailed analysis is necessary before going on, since in the
above fitting the bremsstrahlung and inverse Compton components are fixed:
these components will change following the electron spectrum
which is related to the proton spectrum through the cosmic-ray proton to 
electron ratio, which is usually believed to be universal.

On the other hand, {\sl COS B} data indicate that the diffuse gamma-ray 
spectrum in the inner Galaxy ($310^\circ < \ell < 50^\circ$) is consistent 
with the cosmic-ray proton spectrum of about $\propto E_p^{-2.7}$ 
(Bloemen 1987), which does not seem to be 
consistent with the EGRET result (Hunter et al.\markcite{Hun96} 1996).

Further observations, including analysis of the EGRET data at high latitudes
(Sreekumar et al.\markcite{Sre96} 1996), are nessesary to resolve the 
spectral variation problem.

\section{ Conclusion }

The gamma-ray spectrum from the decay of secondary neutral pions produced by
interactions of cosmic-ray protons with interstellar gas has been calculated
utilizing modern event generators which simulate high-energy $p$-$p$
 collisions by Monte Carlo methods.
The result is not inconsistent with previous calculations: 
the observed excess of the Galactic diffuse gamma-rays above about 1 GeV can
not be explained by revising the models of cosmic-ray interaction with
interstellar matter.
This might suggest flatter cosmic-ray proton spectra in the central region 
of the Galaxy than in the solar neighorhood, but requires further study.

\acknowledgments

I gratefully acknowledge the EGRET team for the kind hospitality 
during my stay at NASA/GSFC, especially Dr.\ Stan Hunter for discussions 
and providing me the EGRET diffuse gamma-ray fluxes prior to publication, and
Dr.\ Robert Hartman for his continuous encouragement and support.
I also thank Drs.\ Jonathan Ormes, Charles Dermer, and Felix Aharonian 
for valuable discussions and suggestions.

\clearpage

\figcaption[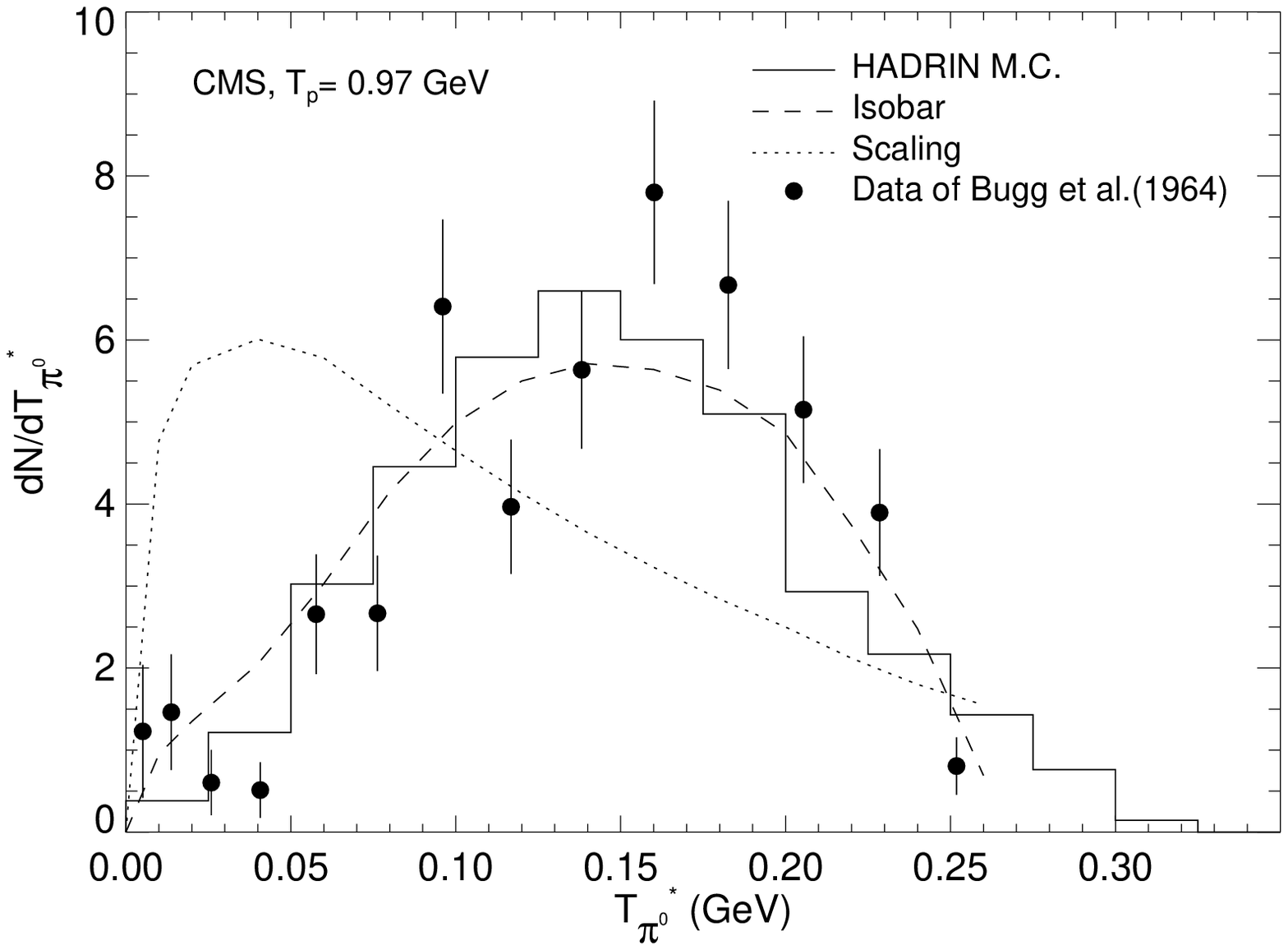]
{Comparison of the model predictions with experimental data of
the secondary $\pi^0$ spectrum in the CMS, corresponding to LS proton kinetic
energy $T_p=0.97$ GeV. Solid histogram: {\sc Hadrin}, dashed line: isobar model by
Stecker\protect\markcite{Ste70} (1970), dotted line: scaling model by 
Stephens and Badhwar\protect\markcite{Ste81} (1981), and
asterisks: data of Bugg et al.\protect\markcite{Bug64} (1964).
\label{fig:tp97cms}
}

\figcaption[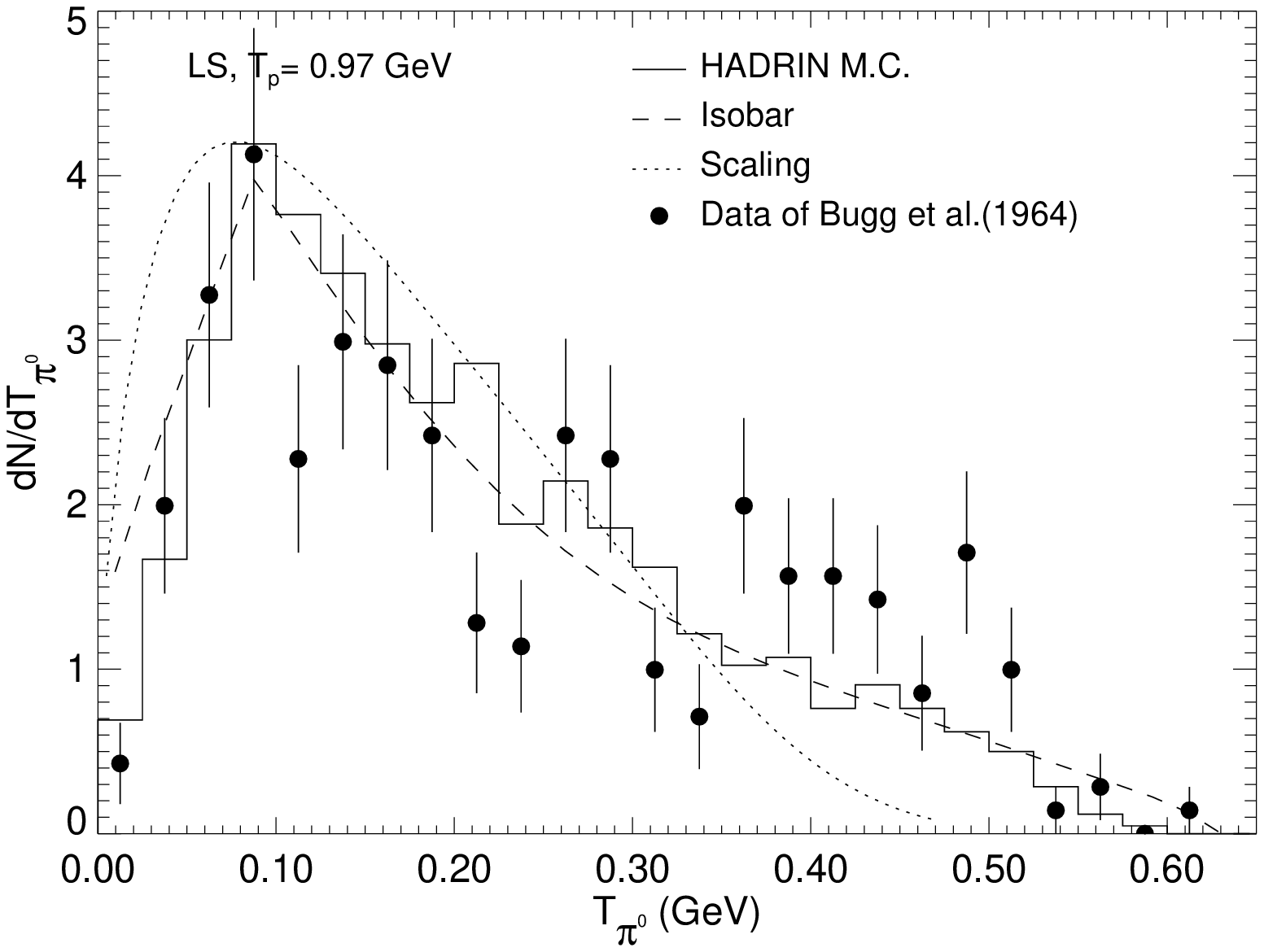]
{Same as in Fig.~\protect\ref{fig:tp97cms}, but with the
secondary $\pi^0$ spectrum in the laboratory system (LS).
\label{fig:tp97ls}
}

\figcaption[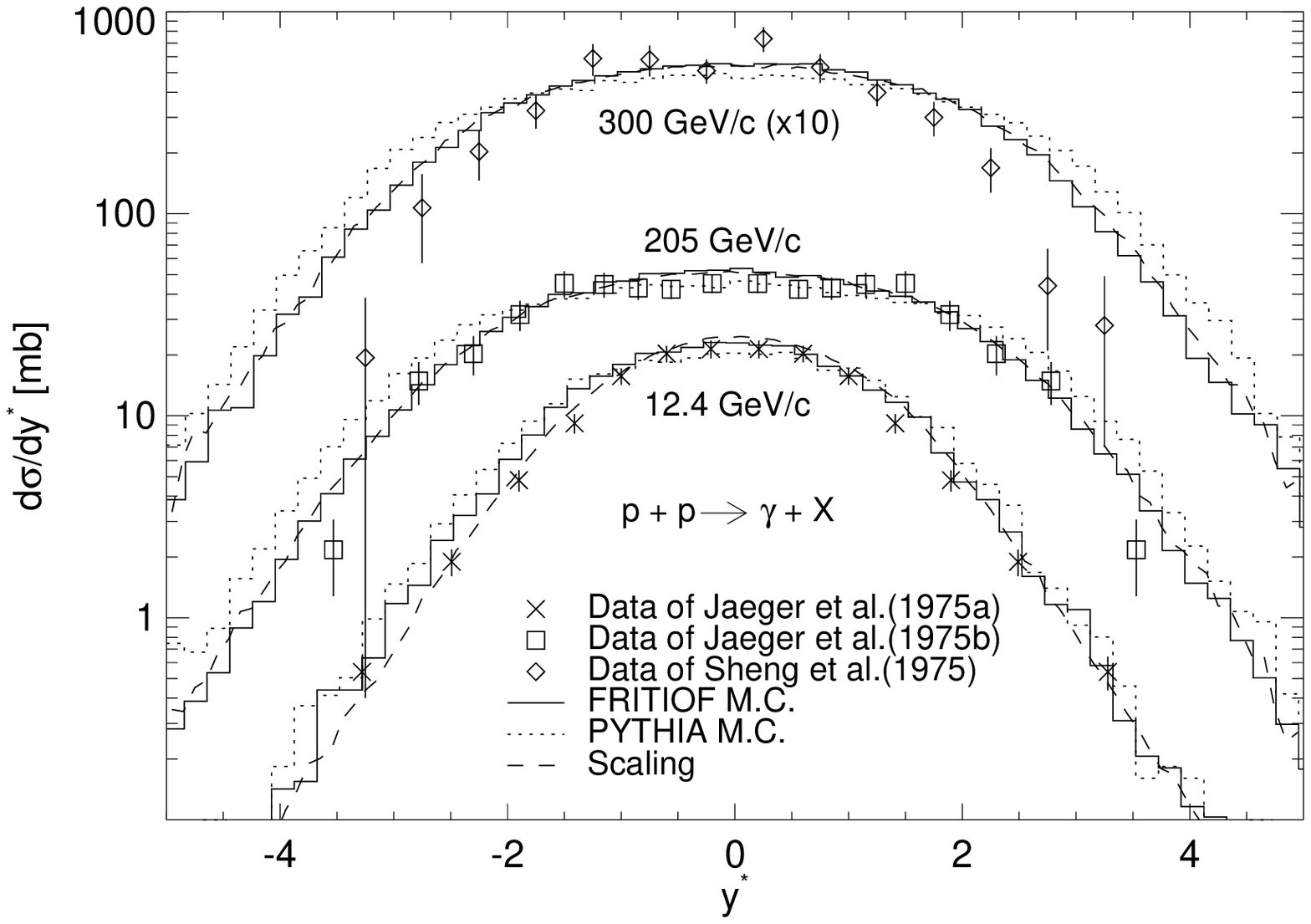]
{Comparison of the model prediction with experimental data
of the rapidity distribution of secondary gamma-rays at LS proton 
momentum of 12.4, 205 and 300 GeV/c.
Solid (dotted) histograms: {\sc Fritiof} ({\sc Pythia}), 
dashed curves: combination of the scaling model
by Stephens \& Badhwar\protect\markcite{Ste81} (1981) and {\sc Decay}.
Calculations are normalized to the inclusive cross section given by
Dermer\protect\markcite{Der86b} (1986b).
Data of Jaeger et al.\protect\markcite{Jae75a} (1975a),
Jaeger et al.\protect\markcite{Jae75b} (1975b) and
Sheng et al.\protect\markcite{She75} (1975).
\label{fig:rapid}
}

\figcaption[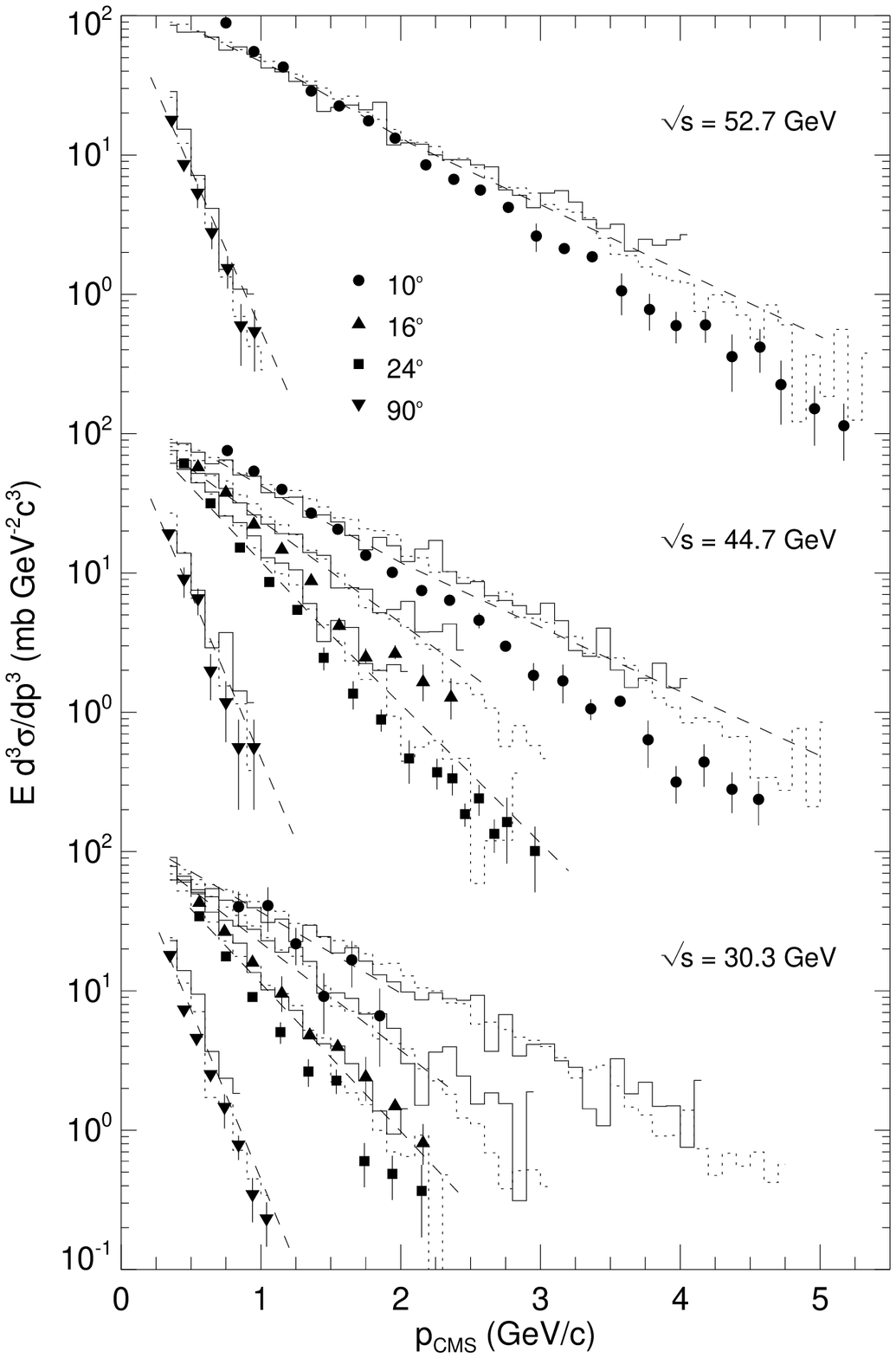]
{Comparison of the invariant cross section for the production
of neutral pions as a function of pion momentum in CMS at different
emission angles for various colliding beam energies.
Data points and scaling calculation (solid lines) are taken from Stephens and 
Badhwar\protect\markcite{Ste81} (1981).
Solid (dotted) histograms are converted from {\sc Fritiof} ({\sc Pythia})
results using the Sternheimer relation (see text).
\label{fig:invcrs}
}

\figcaption[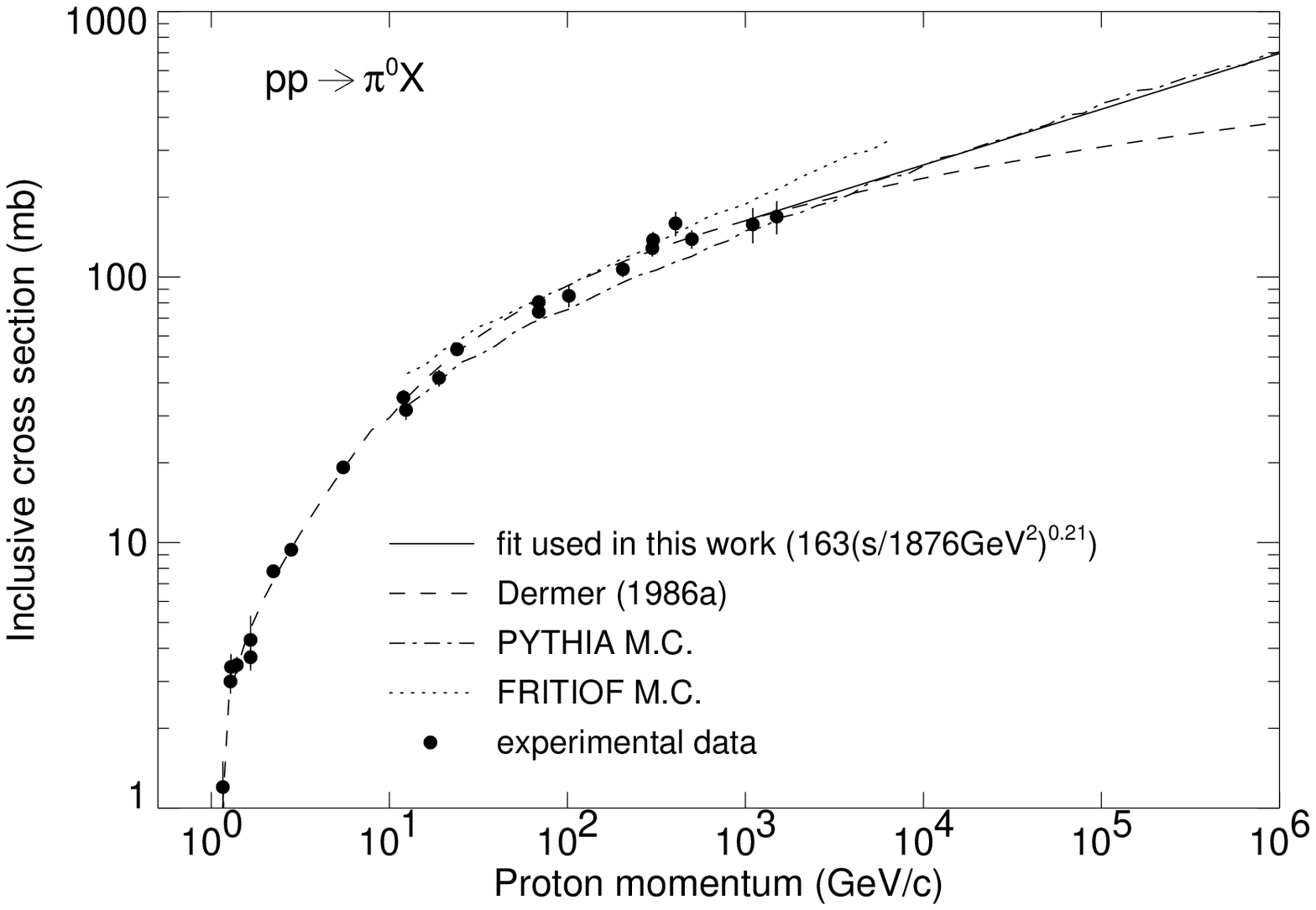]
{Inclusive cross sections for the production of neutral pions in
proton-proton collisions as a function of the incident proton momentum.
The data points are from the compilation of Appendix A of 
Dermer\protect\markcite{Der86b} (1986b) and the dashed lines are the fits by 
Dermer\protect\markcite{Der86b} (1986b). 
The results of {\sc Fritiof} ({\sc Pythia}) are plotted in dotted 
(dotdashed) lines and a fit used in this work is shown by a solid line.
\label{fig:picrs}
}

\figcaption[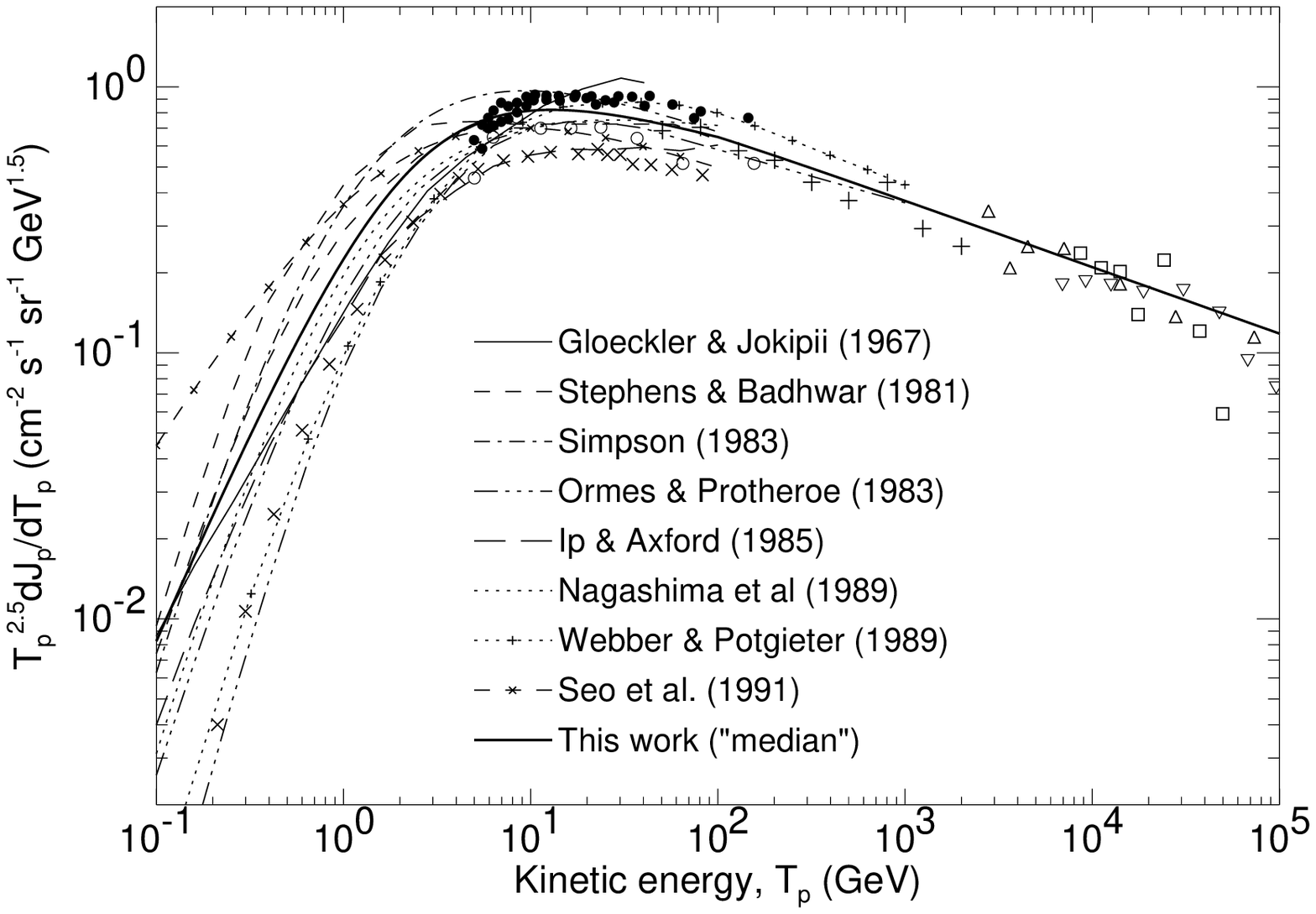]
{Compilation of the demodulated cosmic-ray proton spectra 
calculated by different
authors and the data on the cosmic-ray proton flux at high energies.
Lines are the demodulated 
spectra\protect\markcite{Glo67,Ste81,Sim83,Orm83,Ip85,Nag89,Web89,Seo91} 
(Gloeckler and Jokipii 1967, Stephens and Badhwar 1981, Simpson 1983,
Ormes and Protheroe 1983, Ip and Axford 1985, Nagashima et al.~1989,
Webber and Potgieter 1989, Seo et al.~1991)
and symbols are the data at high energies (pluses, open circles, closed
circles, squares, crosses, inverse triangles and triangles are 
from\protect\markcite{Rya72,Smi73,Web87,Kaw89,Seo91,Asa93,Iva93}
Ryan et al.~1972, Smith et al.~1973, Webber et al.~1987, Kawamura et al.~1989,
Seo et al.~1991, Asakimori et al.~1993 and Ivanenko et al.~1993, respectively.)
Also shown by thick solid line is the ``median'' flux used in this work. See
also Fig.~\protect\ref{fig:crspecMM}. (The upper and lower dashed lines 
correspond to the curves BS and Mu of Stephens and 
Badhwar\protect\markcite{Ste81} (1981).)
\label{fig:crspec}
}

\figcaption[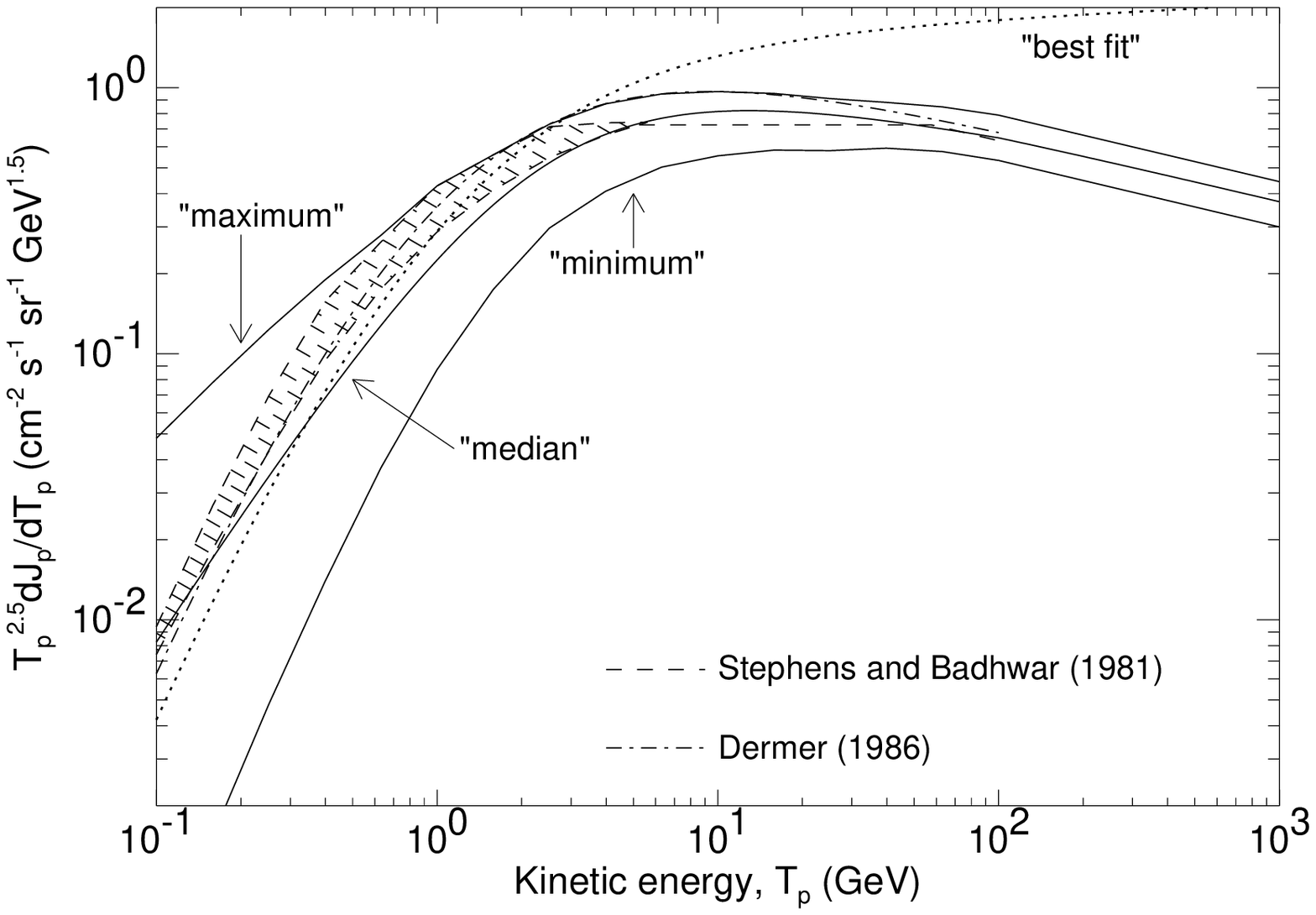]
{The demodulated cosmic-ray proton spectra used in this work:
``median'', ``maximum'', ``minimum'' and ``best fit'', where the ``maximum''
(``minimum'') is taken as the upper (lower) envelopes of different 
calculations and data shown in 
Fig.~\protect\ref{fig:crspec}.
``Best fit'' is derived from the EGRET data (see text).
 Also shown are the spectra used in previous
calculation\protect\markcite{Ste81,Der86a} (Stephens and Badhwar 1981,
Dermer 1986a) of the gamma-ray emissivity.
(The upper and lower dashed lines 
correspond to the curves BS and Mu of Stephens and 
Badhwar\protect\markcite{Ste81} (1981).)
\label{fig:crspecMM}
}

\figcaption[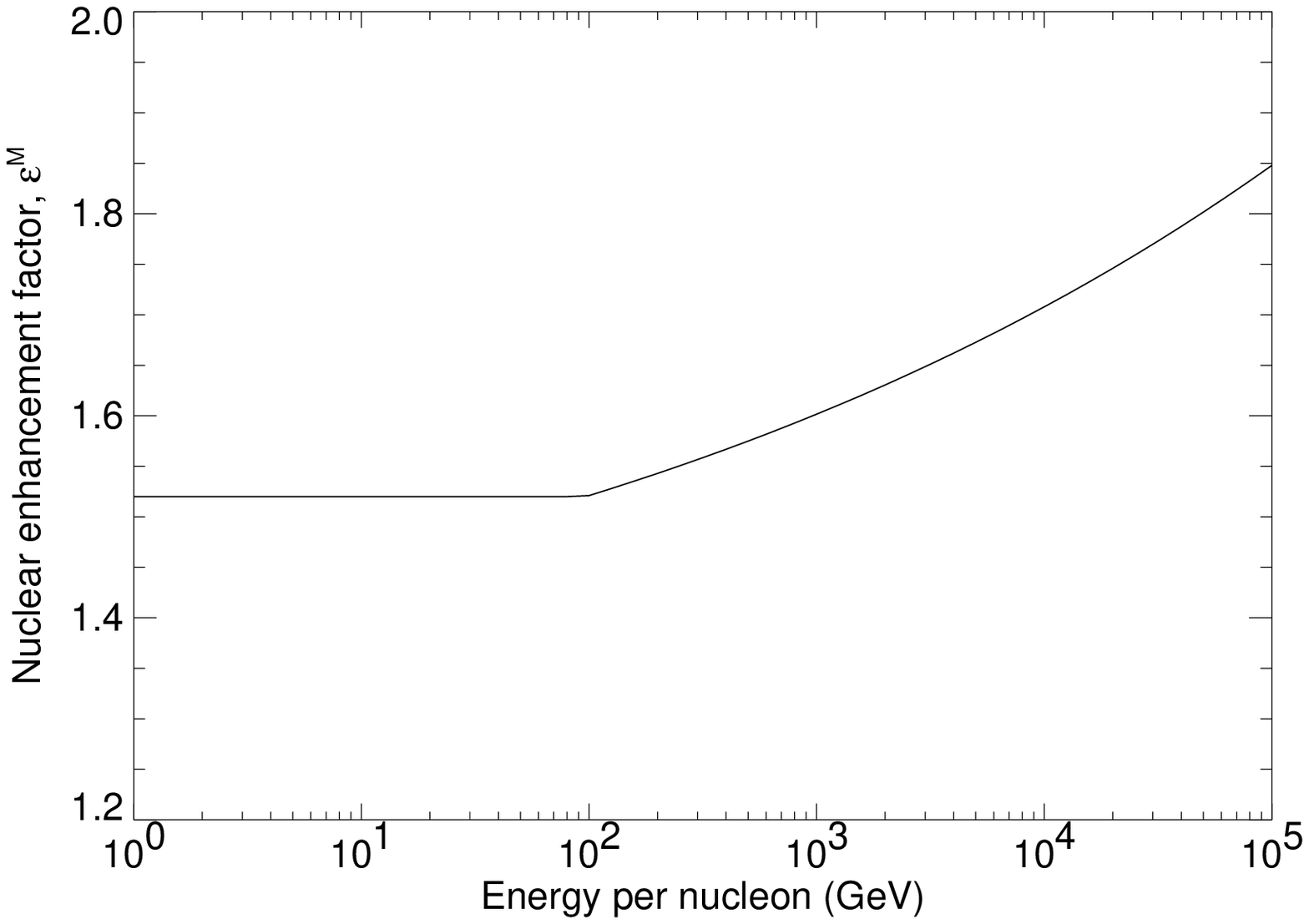]
{The nuclear enhancement factor, $\epsilon^M$, as a function of
the incident cosmic-ray proton energy.
\label{fig:epsM}
}

\figcaption[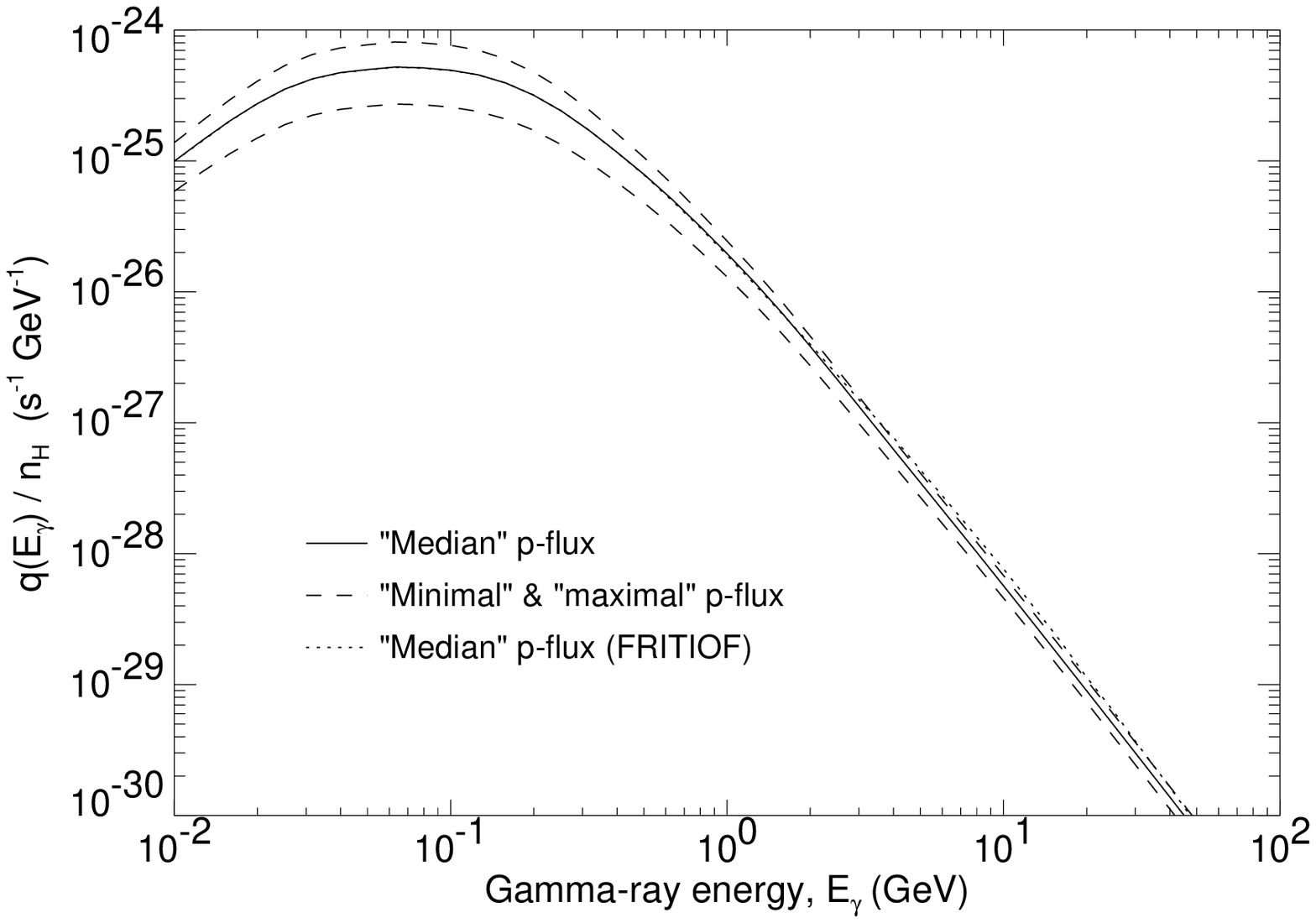]
{The gamma-ray production source function per unit density of 
atomic hydrogen from the interaction of the cosmic-ray with the interstellar 
medium for different models of high-energy $p$-$p$ collisions.
Solid and dotted lines are the results with the scaling model and 
{\sc Fritiof}, respectively, for $T_p>10$ GeV assuming the
``median'' cosmic-ray proton flux.
Both of them used the same model for $T_p\le10$ GeV (see text).
Also shown are the results assuming the ``minimal'' and ``maximal'' proton 
fluxes with the scaling model.
\label{fig:spec_d}
}

\figcaption[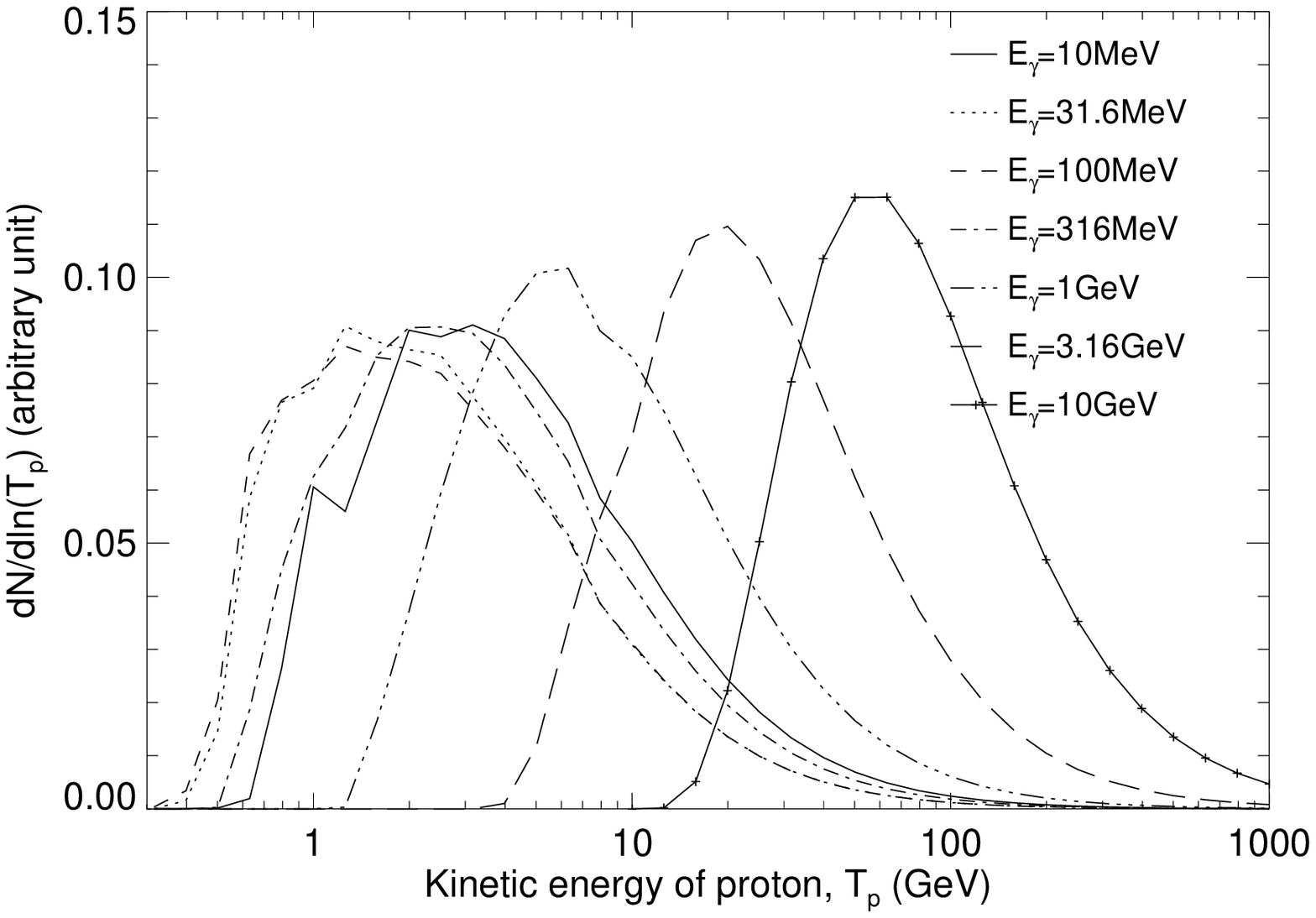]
{The contributions of cosmic-ray protons 
of various energies to gamma-rays of some specific energies.
\label{fig:fixedspec}
}

\figcaption[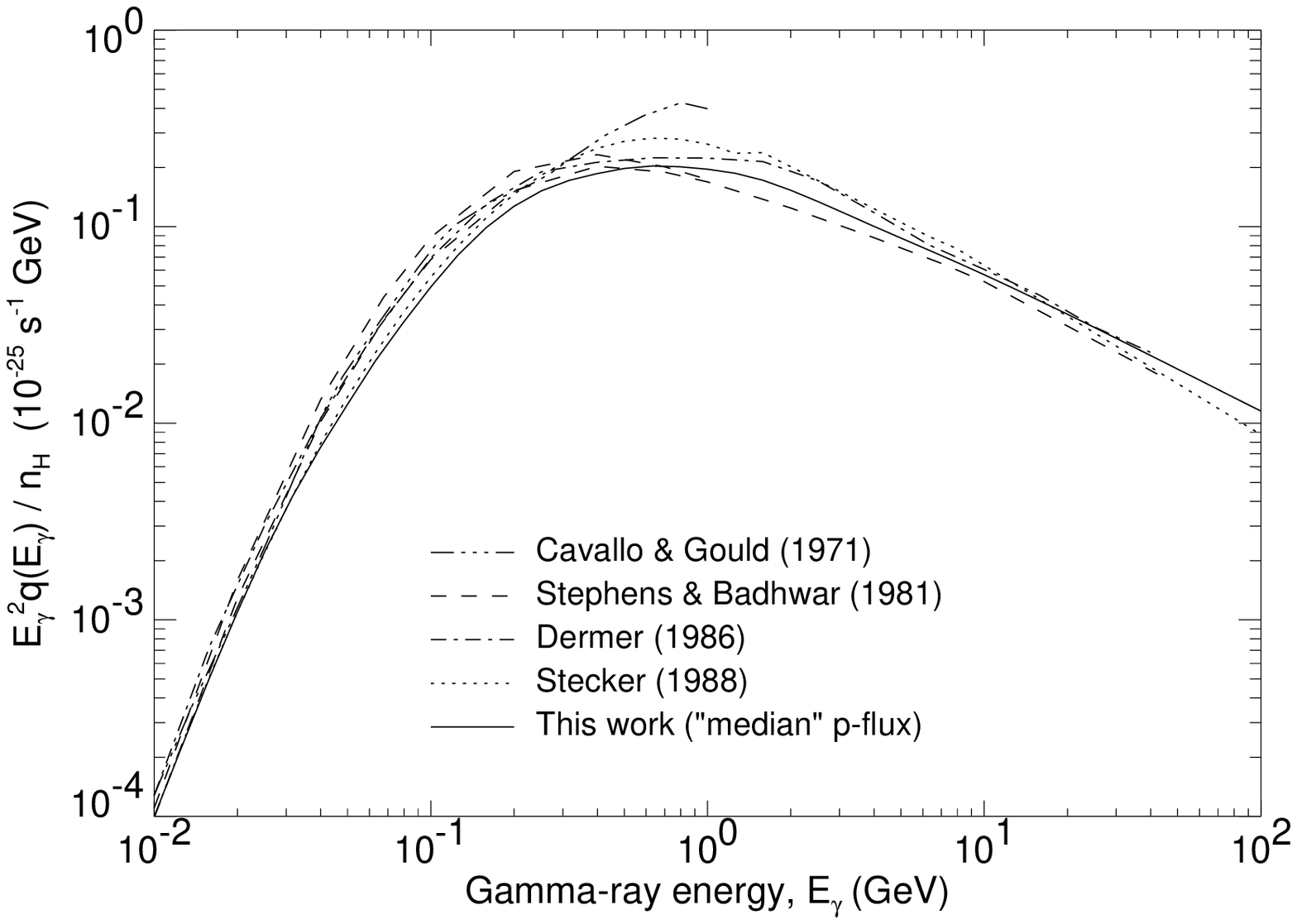]
{Comparison of the gamma-ray production source functions per unit 
density of atomic hydrogen from the interaction of the cosmic-ray with the 
interstellar medium by different authors.
Triple-dot-dashed, dashed, dotdashed lines are differential emissivities
multiplied by $E_\gamma^2$ given in Cavallo and Gould (1971), Stephens and 
Badhwar (1981), Dermer (1986a) and Stecker 
(1988)\protect\markcite{Cav71,Ste81,Der86a,Ste88}, respectively
(the dotted curve is drawn by the parameterization 
given in Bersch et al.\protect\markcite{Ber93} (1993), which fits 
the results of Stecker\protect\markcite{Ste88} (1988)), and the solid
line is from this work with {\sc Pythia} and the ``median'' proton flux.
\label{fig:spec_cmp}
}

\figcaption[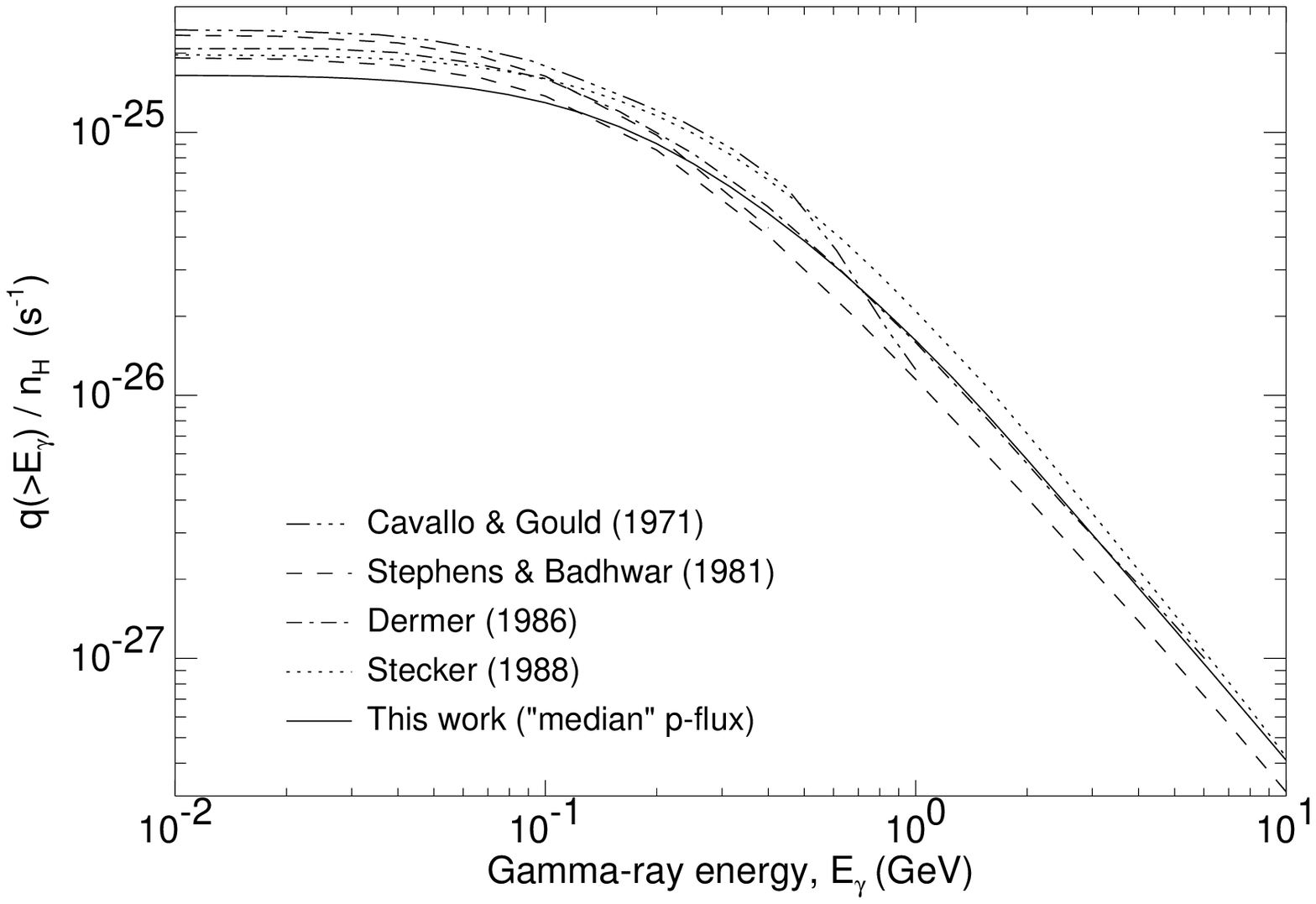]
{Comparison of the integrated gamma-ray production source functions per unit 
density of atomic hydrogen from the interaction of the cosmic-ray with 
the interstellar medium by different authors.
Notations are the same as in Fig.~\protect\ref{fig:spec_d}.
\label{fig:spec_i}
}

\figcaption[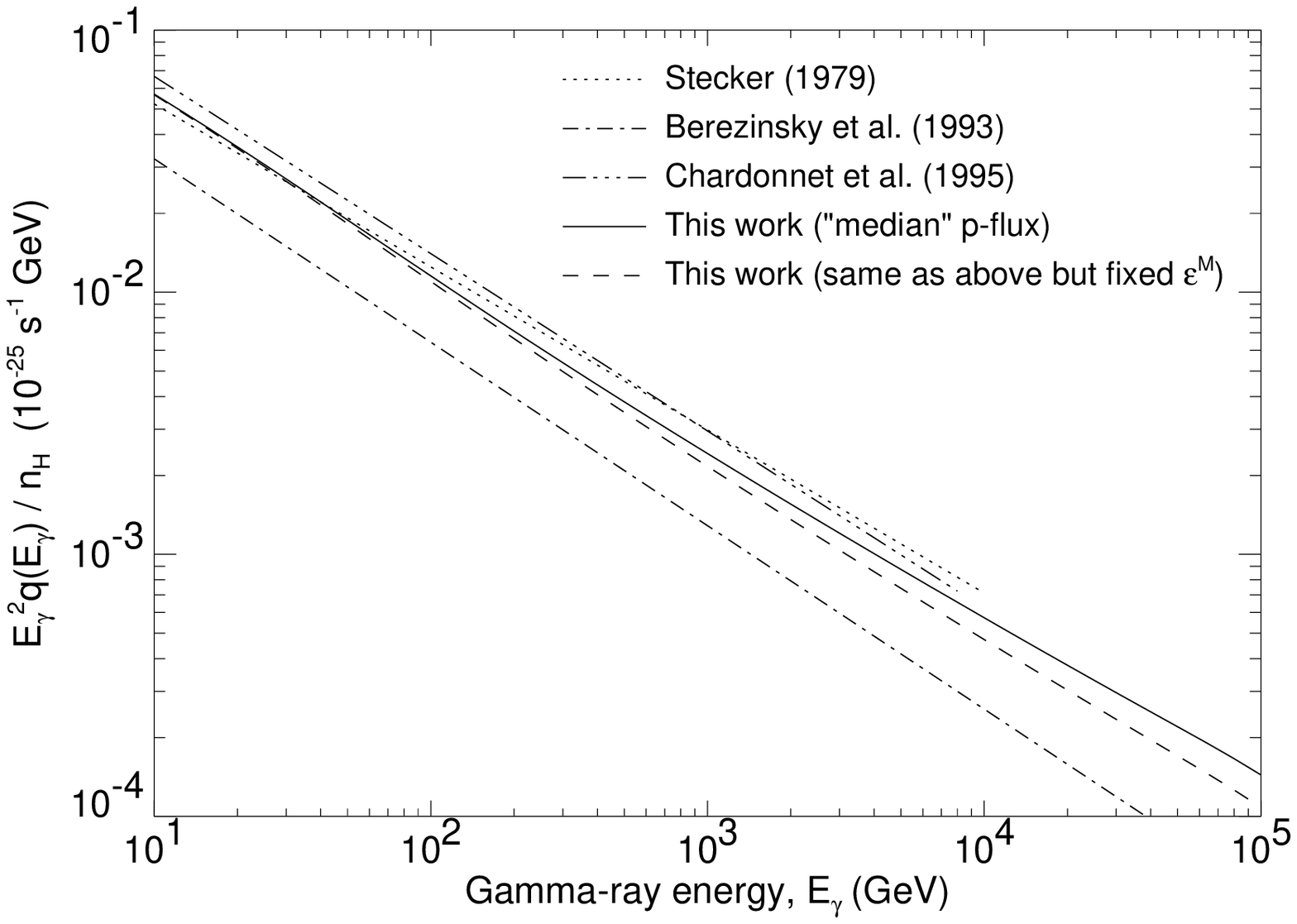]
{The gamma-ray production source functions per unit density of atomic hydrogen
from the interaction of the cosmic-ray with the interstellar medium
for different models of high-energy $p$-$p$ collisions at high energies.
Solid lines are the results with the scaling model and the ``median'' 
proton flux, dashed lines are similar but with a fixed nuclear 
enhancement factor ($\epsilon^M\equiv 1.52$).
Also shown are previous calculations\protect\markcite{Ste79,Ber93a,Cha95}
(Stecker 1979, Berezinsky et al.~1993, Chardonnet et al.~1995).
\label{fig:specHE}
}

\figcaption[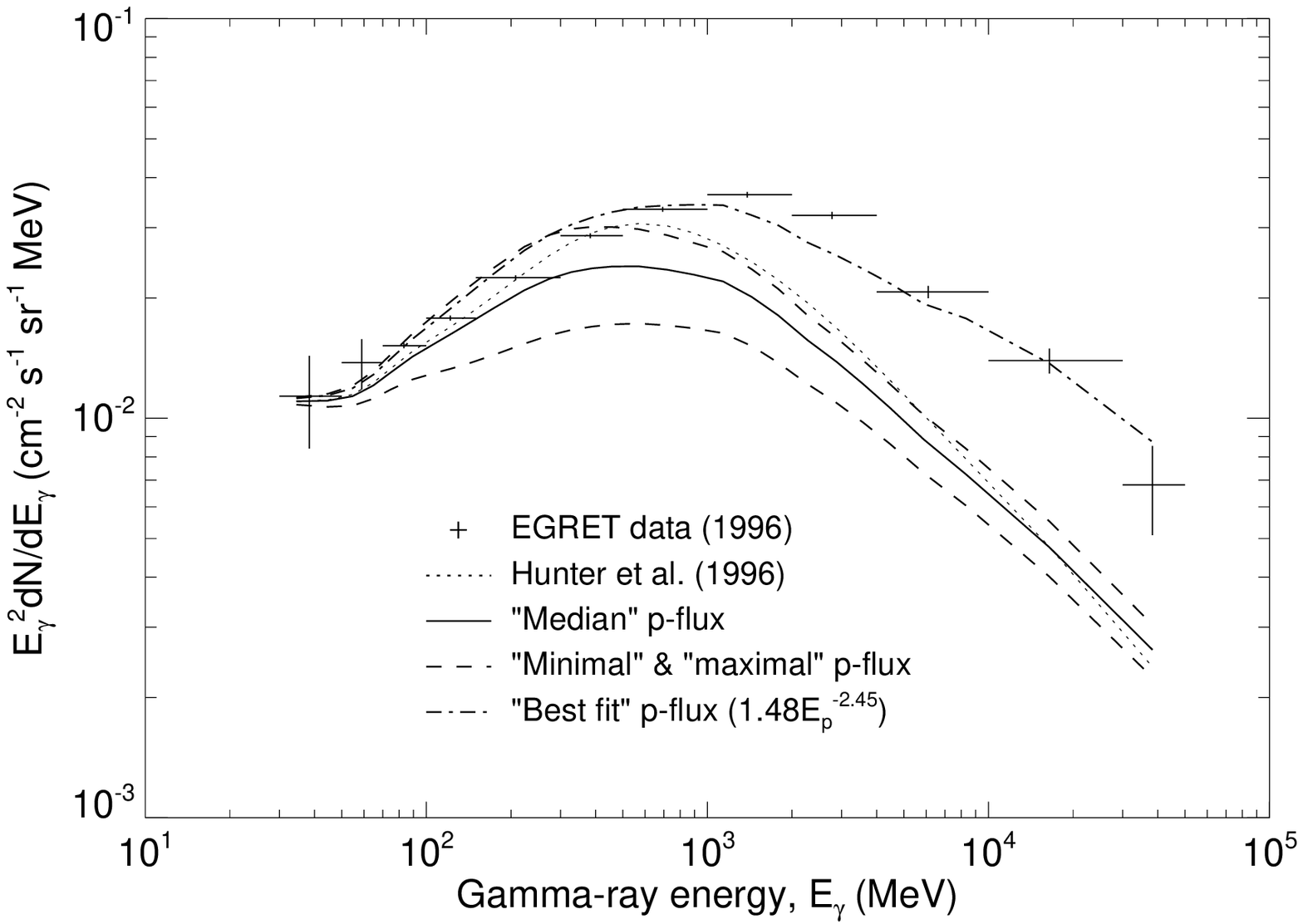]
{The diffuse gamma-ray differential spectrum, multiplied by
$E_\gamma^2$, of the Galactic center region.
Data are from EGRET observations\protect\markcite{Hun96} (Hunter et al.~1996)
averaged over $300^\circ < \ell < 60^\circ$, $|b|\le10^\circ$.
The model predictions based on
gamma-ray emissivities from cosmic-ray nuclear interactions with different
assumptions are also shown. Solid line: the scaling model and the 
``median'' proton flux, dashed lines: the scaling model and the 
``maximum'' and ``minimum'' proton flux, and dotted line: 
from Hunter et al.\protect\markcite{Hun96} (1996) based on the result 
of Stecker\protect\markcite{Ste88} (1988). 
Also shown by dotdashed lines are spectra obtained with the scaling model and
the ``best fit'' proton flux ($\propto E_p^{-2.45}$).
See text for detail.
\label{fig:flux_cmp}
}

\epsscale{0.75}

\clearpage
\plotone{f1.eps}

\noindent{Fig.~\ref{fig:tp97cms}}

\plotone{f2.eps}

\noindent{Fig.~\ref{fig:tp97ls}}

\clearpage
\plotone{f3.eps}

\noindent{Fig.~\ref{fig:rapid}}

\plotone{f4.eps}

\noindent{Fig.~\ref{fig:invcrs}}

\clearpage

\plotone{f5.eps}

\noindent{Fig.~\ref{fig:picrs}}

\plotone{f6.eps}

\noindent{Fig.~\ref{fig:crspec}}

\clearpage
\plotone{f7.eps}

\noindent{Fig.~\ref{fig:crspecMM}}

\plotone{f8.eps}

\noindent{Fig.~\ref{fig:epsM}}

\clearpage
\plotone{f9.eps}

\noindent{Fig.~\ref{fig:spec_d}}

\plotone{f10.eps}

\noindent{Fig.~\ref{fig:fixedspec}}

\clearpage
\plotone{f11.eps}

\noindent{Fig.~\ref{fig:spec_cmp}}

\plotone{f12.eps}

\noindent{Fig.~\ref{fig:spec_i}}

\clearpage
\plotone{f13.eps}

\noindent{Fig.~\ref{fig:specHE}}

\plotone{f14.eps}

\noindent{Fig.~\ref{fig:flux_cmp}}

\end{document}